\RequirePackage{fix-cm} 
\documentclass{svmult}
\usepackage[utf8]{inputenc}
\usepackage{geometry}
\usepackage{natbib}
\usepackage{amssymb}

\usepackage{amsthm}
\usepackage{color}
\usepackage[vlines]{tabularht}
\usepackage{adjustbox}
\usepackage{pdflscape}
\usepackage{array}
\usepackage[table,xcdraw]{xcolor}
\usepackage{multirow}
\usepackage{rotating,tabularx}
\usepackage{url}
\usepackage{tabularx}
\usepackage{mathptmx}
\usepackage{helvet}
\usepackage{courier}
\usepackage{type1cm}        
\usepackage{makeidx}
\usepackage{graphicx}
\usepackage{multicol}
\usepackage[bottom]{footmisc}
\usepackage{psfrag} 
\usepackage{subfigure}
\usepackage{stfloats} 

\makeindex
\DeclareUrlCommand\url{\color{black}\def\UrlLeft{https://}\urlstyle{tt}}

\begin{document}
\pagestyle{plain}
\mainmatter  

\title*{Supervised Learning for the Prediction of Firm Dynamics}

\titlerunning{Supervised learning for the prediction of firm dynamics}

%
%
\author{Falco J. Bargagli-Stoffi \and Jan Niederreiter \and Massimo Riccaboni}
\authorrunning{}

\institute{Falco J. Bargagli-Stoffi \at Harvard University, 677 Huntington Ave, Boston, MA 02115, United States. \\ E-mail: fbargaglistoffi@hsph.harvard.edu.
\and Jan Niederreiter \at IMT School for Advanced studies Lucca, Piazza S. Ponziano, 6 55100 Lucca, Italy. \\ E-mail: jan.niederreiter@alumni.imtlucca.it.
\and Massimo Riccaboni \at IMT School for Advanced studies Lucca, Piazza S. Ponziano, 6 55100 Lucca, Italy. \\E-mail: massimo.riccaboni@imtlucca.it.}

%
%

\toctitle{}
\tocauthor{}
\maketitle
\thispagestyle{empty}

\abstract{Thanks to the increasing availability of granular, yet high-dimensional, firm level data, machine learning (ML) algorithms have been successfully applied to address multiple research questions related to firm dynamics. Especially supervised learning (SL), the branch of ML dealing with the prediction of labelled outcomes, has been used to better predict firms' performance. In this contribution, we will illustrate a series of SL approaches to be used for prediction tasks, relevant at different stages of the company life cycle. The stages we will focus on are (i) startup and innovation, (ii) growth and performance of companies, and (iii) firms exit from the market.
First, we review SL implementations to predict successful startups and R\&D projects. Next, we describe how SL tools can be used to analyze company growth and performance. Finally, we review SL applications to better forecast financial distress and company failure. In the concluding Section, we extend the discussion of SL methods in the light of targeted policies, result interpretability, and causality.}


\section{Introduction}
\noindent In recent years, the ability of machines to solve increasingly more complex tasks has grown exponentially \citep{sejnowski2018deep}. The availability of learning algorithms that deal with tasks such as facial and voice recognition, automatic driving, and fraud detection makes the various applications of machine learning a hot topic not just in the specialised literature but also in the media outlets. Since many decades, computer scientists have been using algorithms that automatically update their course of action to better their performance. Already in the 1950's, Arthur Samuel developed a program to play checkers that improved its performance by learning from its previous moves. The term ``machine learning'' (ML) is often said to have originated in that context. Since then, major technological advances in data storage, data transfer, and data processing have paved the way for learning algorithms to start playing a crucial role in our everyday life.

Nowadays, the usage of ML has become a valuable tool for enterprises' management to predict key performance indicators and thus to support corporate decision-making across the value chain including the appointment of directors \citep{erel2018selecting}, the prediction of product sales \citep{bajari2019impact}, and employees' turnover \citep{ajit2016prediction,saradhi2011employee}. Using data which emerges as a byproduct of economic activity has a positive impact on firms' growth \citep{farboodi2019big} and strong data analytic capabilities leverage corporate performance \citep{mikalef2019big}. Simultaneously, publicly accessible data sources that cover information across firms, industries and countries, open the door for analysts and policy makers to study firm dynamics on a broader scale such as the fate of start-ups \citep{guerzoni2019survival}, product success \citep{munos2019}, firm growth \citep{weinblat2018forecasting}, and bankruptcy \citep{bargagli2020machine}.

Most ML methods can be divided in two main branches: (i) \textit{unsupervised learning} (UL) and (ii) \textit{supervised learning} (SL) models. 
UL refers to those techniques used to draw inferences from data sets consisting of input data without labelled responses. These algorithms are used to perform tasks such as clustering and pattern mining.
SL refers to the class of algorithms employed to make predictions on labelled response values (i.e., discrete and continuous outcomes). In particular, SL methods use a known data set with input data and response values, referred as training data set, to learn how to successfully perform predictions on the labelled outcomes. The learned decision rules can then be used to predict unknown outcomes of new observations. For example, an SL algorithm could be trained on a data set that contains firm-level financial accounts and information on enterprises solvency status in order to develop decision rules that predict the solvency of companies.

SL algorithms provide great added value in predictive tasks since they are specifically designed for such purposes \citep{kleinberg2015prediction}. Moreover, the non-parametric nature of SL algorithms makes them suited to uncover hidden relationships between the predictors and the response variable in large data sets that would be missed out by traditional econometric approaches. Indeed, the latter models, e.g, ordinary least squares and logistic regression, are built assuming a set of restrictions on the functional form of the model to guarantee statistical properties such as estimator unbiasedness and consistency. 
SL algorithms often relax those assumptions and the functional form is dictated by the data at hand (data-driven models). This characteristic makes SL algorithms more ``adaptive" and inductive, therefore enabling more accurate predictions for future outcome realizations.

In this contribution, we focus on the traditional usage of SL for predictive tasks, excluding from our perspective the growing literature that regards the usage of SL for causal inference. As argued by \cite{kleinberg2015prediction}, researchers need to answer to both causal and predictive questions in order to inform policy makers. An example that helps us to draw the distinction between the two is provided by a policy maker facing a pandemic.
On the one side, if she wants to assess whether a quarantine will prevent a pandemic to spread, she needs to answer a purely causal question (i.e., ``what is the effect of a quarantine on the chance to that the pandemic will spread?"). On the other side, if she wants to know if she should start a vaccination campaign, she needs to answer to a purely predictive question (i.e., ``is the pandemic going to spread within the country?''). SL tools can help policy makers to navigate both these sorts of policy relevant questions \citep{mullainathan2017machine}. We refer to \cite{athey2019machine} and \cite{athey2018impact} for a critical review of the causal machine learning literature.

Before getting into the nuts and bolts, we want to highlight that our goal is not to provide a comprehensive review of all the applications of SL for prediction of firm dynamics, but to describe the alternative methods used so far in this field. Namely, we selected papers based on the following inclusion criteria: (i) the usage of SL algorithm to perform a predictive task in one of the fields of our interest (i.e., enterprises success, growth or exit); (ii) a clear definition of the outcome of the model and the predictors used; (iii) an assessment of the quality of the prediction.
The purpose of this work is twofold. First, we outline a general SL framework to ready the readers' mindset to think about prediction problems from an SL-perspective (Section \ref{sec:sl}). Second, equipped with the general concepts of SL, we turn to real-world applications of the SL predictive power in the field of firms' dynamics. 
Due to the broad range of SL-applications, we organize Section \ref{sec:firm_dynamics} into three parts according to different stages of the firm lifecycle. The prediction tasks we will focus on are about the success of new enterprises and innovation (Section \ref{subsec:success}), firm performance and growth (Section \ref{subsec:performance}), and the exit of established firms (Section \ref{subsec:exit}). The last Section discusses the state-of-the-art, future trends and relevant policy implications (Section \ref{sec:discussion}). 

\section{Supervised machine learning} \label{sec:sl}

In a famous paper on the difference between model-based and data-driven statistical methodologies Berkeley Professor Leo Breiman stated, referring to the statistical community, that ``\textit{there are two cultures in the use of statistical modeling to reach conclusions
from data. One assumes that the data are generated by a given stochastic data model. The other uses algorithmic models and treats the data mechanism as unknown.} [...] \textit{If our goal as a field is to use data to solve problems, then we need to move away from exclusive dependence on data models and adopt a diverse set of tools}"  \citep[][p. 199]{breiman2001statistical}.
In this quote Breiman catches the essence of SL algorithms: their ability to capture hidden patterns in the data by directly learning from them, without the restrictions and assumptions of model-based statistical methods.

SL algorithms employ a set of data with input data and response values, referred as training sample, to learn and make predictions (in-sample predictions) while another set of data, referred as test sample, is kept separate to validate the predictions (out-of-sample predictions). Training and testing sets are usually built by randomly sampling observations from the initial data set. In the case of panel data, the testing sample should contain only observations that occurred later in time than the observations used to train the algorithm to avoid the so-called \textit{look-ahead bias}. This ensures that future observations are predicted from past information, not vice versa.

When the dependent variable is categorical (e.g., yes/no or category 1--5) the task of the SL algorithm is referred as a ``classification" problem, whereas in ``regression" problems the dependent variable is continuous.

The common denominator of SL algorithms is that they take an information set $\mathbf{X}_{N \times P}$, i.e. a matrix of features (also referred to as attributes or predictors), and map it to an $N$-dimensional vector of outputs $y$ (also referred to as actual values or dependent variable), where $N$ is the number of observations $i=1,\ldots,N$ and $P$ is the number of features. The functional form of this relationship is very flexible and gets updated by evaluating a loss function. The functional form is usually modelled in two steps \citep{mullainathan2017machine}:
\begin{enumerate}
        \item pick the best in-sample loss-minimizing function $f(\cdot)$:
		\begin{equation} \label{MLequation} \footnotesize
		argmin \sum_{i=1}^{N} L\big(f(x_i), y_i\big) \: \: \: \:  over  \: \:  \: \: f(\cdot) \in F \: \: \: \:  \: \: \: \: s. \: t.\: \:  \: \: \: \:  \: \: R\big(f(\cdot)\big) \leq c
		\end{equation}
		where $\sum_{i=1}^{N} L\big(f(x_i), y_i\big)$ is the in-sample loss functional to be minimized (i.e., the mean-squared-error of prediction), $f(x_i)$ are the predicted (or fitted) values, $y_i$ are the actual values, $f(\cdot) \in F$ is the function class of the SL algorithm and $R\big(f(\cdot)\big)$ is the complexity functional that is constrained to be less than a certain value $c \in \mathbb{R}$ (e.g., one can think of this parameter as a budget constraint);
		\item estimate the optimal level of complexity using empirical tuning through cross-validation.
\end{enumerate}
    
Cross-validation refers to the technique that is used to evaluate predictive models by training them on the training sample, and evaluating their performance on the test sample.\footnote{This technique (hold-out) can be extended from two to $k$ folds. In $k$-folds cross-validation, the original data set is randomly partitioned into $k$ different subsets. The model is constructed on $k-1$ folds and evaluated on 1 fold repeating the procedure until all the $k$ folds are used to evaluate the predictions.}
 Then, on the test sample the algorithm's performance is evaluated on how well it has learned to predict the dependent variable $y$. 
 By construction, many SL algorithms tend to perform extremely well on the training data. This phenomenon is commonly referred as ``over-fitting the training data" because it combines very high predictive power on the training data with poor fit on the test data. This lack of generalizability of the model's prediction from one sample to another can be addressed by penalizing the model's complexity. The choice of a good penalization algorithm is crucial for every SL technique to avoid this class of problems.
 
\begin{figure}[!htbp]

        \caption{Exemplary confusion matrix for assessment of classification performance}
         \label{fig:confusionmatrix}
        \includegraphics[width=114mm]{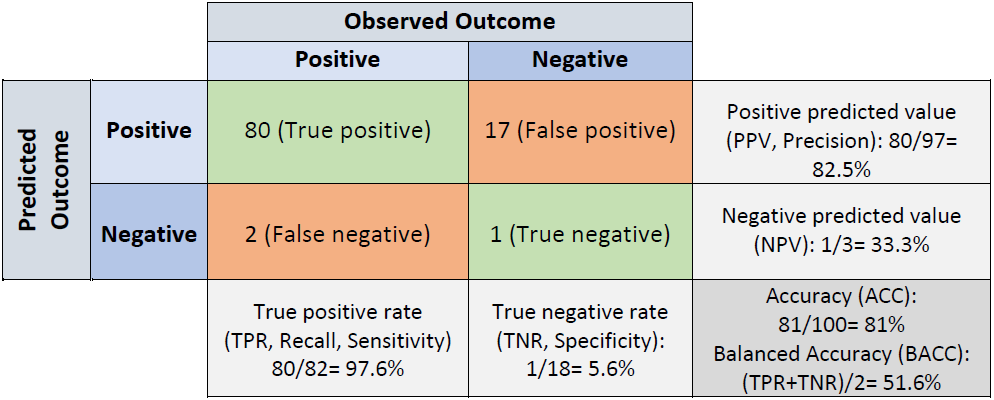}\\
\end{figure}

 In order to optimize the complexity of the model, the performance of the SL algorithm can be assessed by employing various performance measures on the test sample.
 It is important for practitioners to choose the performance measure that fits best the prediction task at hand and the structure of the response variable.
 In regression tasks different performance measures can be employed. The most common ones are the mean-squared-error (MSE), the mean-absolute-error (MAE) and the $R^2$.
 In classification tasks the most straightforward method is to compare true outcomes with predicted ones via confusion matrices from where common evaluation metrics, such as true positive rate (TPR), true negative rate (TNR), and accuracy (ACC), can be easily calculated (see Figure \ref{fig:confusionmatrix}).
Another popular measure of prediction quality for binary classification tasks (i.e., positive vs. negative response), is the Area Under the receiver operating Curve (AUC) that relates how well the trade-off between the models TPR and TNR is solved. TPR refers to the proportion of positive cases that are predicted correctly by the model, while TNR refers to the proportion of negative cases that are predicted correctly. Values of AUC range between 0 and 1 (perfect prediction), where 0.5 indicates that the model has the same prediction power as a random assignment. The choice of the appropriate performance measure is key to communicate the fit of a SL model in an informative way. 

 Consider the example in Figure \ref{fig:confusionmatrix} in which the testing data contains 82 positive outcomes (e.g. firm survival) and 18 negative outcomes, such as firm exit, and the algorithm predicts 80 of the positive outcomes correctly but only one of the negative ones. The simple accuracy measure would indicate 81\% correct classifications, but the results suggest that the algorithm has not successfully learned how to detect negative outcomes. In such case, a measure that considers the unbalance of outcomes in the testing set, such as balanced accuracy (BACC, defined as ($(TPR+TNR/2)=51.6\%$), or the F1-score would be more suited.
 Once the algorithm has been successfully trained and its out of sample performance has been properly tested, its decision rules can be applied to predict the outcome of new observations, for which outcome information is not (yet) known. 
 
 Choosing a specific SL algorithm is crucial since performance, complexity, computational scalability, and interpretability differ widely across available implementations. In this context, easily interpretable algorithms are those that provide comprehensive decision rules from which a user can retrace results \citep{lee2019machine}.
 Usually highly complex algorithms require the discretionary fine tuning of some model hyper-parameters, more computational resources and their decision criteria are less straightforward. Yet, the most complex algorithms do not necessarily deliver the best predictions across applications \citep{kotthoff2016algorithm}. Therefore, practitioners usually run a \textit{horse race} on multiple algorithms and choose the one that provides the best balance between interpretability and performance on the task at hand. In some learning applications for which prediction is the sole purpose, different algorithms are combined and the contribution of each chosen so that overall predictive performance gets maximized. Learning algorithms that are formed by multiple self-contained methods are called ensemble learners (e.g., the super-learner algorithm by \cite{van2007super}). 
 
 Moreover, SL algorithms are used by scholars and practitioners to perform predictors selection in high-dimensional settings (e.g., scenarios where the number of predictors is larger than the number of observations: small $N$ large $P$ settings), text analytics, and natural language processing (NLP). The most widely used algorithms to perform the former task are the Least-Absolute-Shrinkage-and-Selection-Operator (Lasso) algorithm \citep{tibshirani1996regression} and its related versions such as stability selection \citep{meinshausen2010stability} and C-Lasso \citep{su2016identifying}. The most popular supervised NLP and text analytics SL algorithms are support vector machines \citep{steinwart2008support}, Naive Bayes \citep{ng2002discriminative}, and Artificial Neural Networks (ANN) \citep{hassoun1995fundamentals}.
 
  Reviewing SL algorithms and their properties in detail would go beyond the scope of this contribution, however, in Table \ref{tab:sl_algorithms} we provide a basic intuition of the most widely used SL methodologies employed in the field of firm dynamics. A more detailed discussion of the selected techniques, together with a code example to implement each one of them in the statistical software \texttt{R} and a toy application on real firm-level data, is provided in the following web page:\\ \begin{footnotesize}\url{github.com/fbargaglistoffi/machine-learning-firm-dynamics}\end{footnotesize}
 
  \begin{table}[htbp!]
    \centering
       \caption{SL algorithms commonly applied in predicting firm dynamics}
    \label{tab:sl_algorithms}
    \begin{footnotesize}
    \begin{tabularx}{\textwidth}{>{\raggedright\arraybackslash}p{1.1cm}>{\raggedright\arraybackslash}p{8.4cm}>{\centering\arraybackslash}p{1.6cm}}
    
    \hline\noalign{\smallskip}
             \textbf{Method} &  \textbf{Description} & \textbf{Interpretability}  \\ 
            \noalign{\smallskip}\svhline\noalign{\smallskip}
           Decision Tree (DT) & Decision trees (DT) consist of a sequence of binary decision rules (nodes) on which the tree splits into branches (edges). At each final branch (leaf node) a decision regarding the outcome is estimated. The sequence and definition of nodes is based on minimizing a measure of node purity (e.g., Gini index, or entropy for classification tasks and MSE for regression tasks). Decision trees are easy to interpret but sensitive to changes in the features, that frequently lower their predictive performance \citep[see also][]{breiman2017classification} .& High \tabularnewline
           Random Forest (RF) & Instead of estimating just one DT, random forest (RF) re-samples the training set observations to estimate multiple trees. For each tree at each node a set of $m$ (with $m < P$) predictors is chosen randomly from the features space. To obtain the final prediction the outcomes of all trees are averaged or in classification tasks chosen by majority vote \citep[see also][]{breiman2001random}. & Medium \tabularnewline
           Support vector machines (SVM) & Support vector machine algorithms (SVM) estimate a hyperplane over the feature space to classify observations. The vectors that span the hyperplane are called support vectors. They are chosen such that the overall distance (referred to as margin) between the data points and the hyperplane as well as the prediction accuracy is maximized \citep[see also][]{steinwart2008support}.
            & Medium \tabularnewline
           (Deep) Artificial Neural Networks (ANN) & Inspired by biological networks, every artificial neural network (ANN) consists of, at least, three layers (deep ANNs are ANNs with more than three layers): an input layer with feature information, one or more hidden layer, and an output layer returning the predicted values. Each layer consists of nodes (neurons) that are connected via edges across layers. During the learning process, edges that are more important are reinforced. Neurons may then only send a signal if the signal received is strong enough \citep[see also][]{hassoun1995fundamentals}. & Low \\ 

      \noalign{\smallskip}\hline\noalign{\smallskip}
    \end{tabularx}
   
 \end{footnotesize}
\end{table}

\section{SL prediction of firm dynamics} \label{sec:firm_dynamics}

Here, we review SL applications that have leveraged inter firm data to predict various company dynamics. Due to the increasing volume of scientific contributions that employ SL for company related prediction tasks, we split the Section in three parts according to the life cycle of a firm. In Section \ref{subsec:success} we review SL applications that deal with early stage firm success and innovation, in Section \ref{subsec:performance} we discuss growth and firm performance related work, and lastly, in Section \ref{subsec:exit}, we turn to firm exit prediction problems.
\subsection{Entrepreneurship and innovation} \label{subsec:success}

The success of young firms (referred to as startups) plays a crucial role in our economy since these firms often act as net creator of new jobs \citep{henrekson2010gazelles} and push, through their product and process innovations, the societal frontier of technology. Success stories of Schumpeterian entrepreneurs that reshaped entire industries are very salient, yet from a probabilistic point of view it is estimated that only 10\% of startups stay in business long-term \citep{griffith2014, krishna2016predicting}.

Not only is startup success highly uncertain, but it also escapes our ability to identify the factors to predict successful ventures. Numerous contributions have used traditional regression based approaches to identify factors associated with the success of small businesses \cite[e.g.,][]{lussier2001crossnational,lussier2010three,halabi2014model}, yet do not test the predictive quality of their methods out-of-sample and rely on data specifically collected for the research purpose. 
Fortunately, open access platforms such as \textit{Chrunchbase.com} and \textit{Kickstarter.com} provide company and project specific data whose high dimensionality can be exploited using predictive models \citep{dalle2017using}. SL algorithms, trained on a large amount of data, are generally suited to predict startup success, especially because success factors are commonly unknown and their interactions complex.  Similarly to the prediction of success at the firm level, SL algorithms can be used to predict success for singular projects. Moreover, unstructured data, e.g. business plans, can be combined with structured data to better predict the odds of success.

Table \ref{tab:SL_earlysuccess} summarizes the characteristics of recent contributions in various disciplines that use SL algorithms to predict startup success (upper half of the Table) and success on the project level (lower half of the Table).  The definition of success varies across these contributions. Some authors define successful startups as firms that receive a significant source of external funding (this can be additional financing via venture capitalists, an initial public offering or a buyout) that would allow to scale operations \citep{arroyo2019assessment,bento2018predicting, sharchilev2018web,xiang2012supervised,zhang2017predicting}. Other authors define successful startups as companies that simply survive \citep{bohm2017business,krishna2016predicting, mckenzie2017man} or coin success in terms of innovative capabilities \citep{kinne2019predicting, guerzoni2019survival}. As data on the project level is usually not publicly available \citep{janssen2019machine,dellermann2017finding}, research has mainly focused on two areas for which it is, namely the project funding success of crowdfunding campaigns \citep{etter2013launch,greenberg2013crowdfunding,kaminski2019predicting} and the success of pharmaceutical projects to pass clinical trials \citep{dimasi2015tool,feijoo2020key,lo2019machine,munos2019}.\footnote{Since 2007 the US Food and Drug Administration (FDA) requires that the outcome of clinical trials that passed ``Phase I'' are publicly disclosed \citep{ zarin2016trial}. Information on these clinical trials, and pharmaceutical companies in general, has since then been used to train SL methods to classify the outcome of R\&D projects.}

To successfully distinguish how to classify successes from failures, algorithms are usually fed with company, founder, and investor specific inputs that can range from a handful of attributes to a couple of hundred. Most authors find the information that relate to the source of funds predictive for startup success \citep[e.g.,][]{bento2018predicting, krishna2016predicting, sharchilev2018web} but also entrepreneurial characteristics \citep{mckenzie2017man} and the engagement in social networks \citep{zhang2017predicting} seem to matter. At the project level, funding success depends on the number of investors \citep{greenberg2013crowdfunding} as well on the audio/visual content provided by the owner to pitch the project \citep{kaminski2019predicting}, whereas success in R\&D projects depends on an interplay between company, market and product driven factors \citep{munos2019}.

Yet, it remains challenging to generalize early stage success factors, as these accomplishments are often context dependent and achieved differently across heterogeneous firms. To address this heterogeneity, one approach would be to first categorize firms and then train SL algorithms for the different categories. One can manually define these categories (i.e., country, size cluster, etc.) or adopt a data driven approach \citep[e.g.,][]{su2016identifying}.

The SL methods that best predict startup and project success vary vastly across reviewed applications with random forest (RF) and support vector machine (SVM) being the most commonly used approaches. Both methods are easily implemented (see our web appendix) and despite their complexity still deliver interpretable results, including insights on the importance of singular attributes.  In some applications, easily interpretable logistic regressions (LR) perform at par or better than more complex methods \citep{fantazzini2009random,kaminski2019predicting, krishna2016predicting}.  This might first seem surprising, yet it largely depends on whether complex interdependencies in the explanatory attributes are present in the data at hand. As discussed in Section \ref{sec:sl} it is therefore recommendable to run a horse race to explore the prediction power of multiple algorithms that vary in terms of their interpretability.

Lastly, even if most contributions report their goodness of fit (GOF) using standard measures such as ACC and AUC, one needs to be cautions when cross comparing results because these measures depend on the underlying data set characteristics which may vary. Some applications use data samples, in which successes are less frequently observed than failures. Algorithms that perform well when identifying failures but have limited power when it comes to classifying successes would then be better ranked in terms of ACC and AUC than algorithms for which the opposite holds (see Section \ref{sec:sl}). The GOF across applications simply reflects that SL methods, on average, are useful for predicting startup and project outcomes. However, there is still considerable room for improvement that could potentially  come from the quality of the used features as we do not find a meaningful correlation between data set size and GOF in the reviewed sample.

\subsection{Firm performance and growth}

Despite recent progress \citep{buldyrev2020rise} firm growth is still an elusive problem. Since the seminal contribution of \citet{gibrat1931} firm growth is still considered, at least partially, as a random walk \citep{coad2013growth}, there has been little progress in identifying the main drivers of firm growth \citep{coad2009growth} and recent empirical models have a small predictive power \citep{van2019firm}. Moreover, firms have been found to be persistently heterogeneous with results varying depending on their life stage and marked differences across industries and countries. Although a set of stylized facts are well established, such as the negative dependency of growth on firm age and size, it is difficult to predict the growth and performance from previous information such as balance sheet data. - i.e., it remains unclear what are good predictors for what type of firm.

SL excels at using high dimensional inputs, including non-conventional unstructured information such as textual data, and use them all as predictive inputs. Recent examples from the literature reveal a tendency in using multiple SL tools to make better predictions out of publicly available data sources such as financial reports \citep{qiu2014supervised}, and company web pages \citep{kolkman2019data}. The main goal is to identify the key drivers of superior firm performance in terms of profits, growth rates and return on investments. This is particularly relevant for stakeholders, including investors and policy makers, to devise better strategies for sustainable competitive advantage. For example, one of the objectives of the European commission is to incentivize high growth firms (HGFs) \citep{european2010europe}, which could get facilitated by classifying such companies adequately.

A prototypical example of application of SL methods to predict HGFs is \citet{weinblat2018forecasting} who uses an RF algorithm trained on firm characteristics for different EU countries. He finds that HGFs have usually experienced prior accelerated growth and should not be confused with startups that are generally younger and smaller. Predictive performance varies substantially across country samples, suggesting that the applicability of SL approaches cannot be generalized. 
Similarly, \citet{miyakawa2017forecasting} show that RF can outperform traditional credit score methods to predict firm exit, growth in sales and profits of a large sample of Japanese firms. Even if the reviewed SL literature on firms' growth and performance has introduced approaches that increment predictive performance compared to traditional forecasting methods, it should be noted that this performance stays relatively low across applications in the firms’ life cycle and does not seem to correlate significantly with the size of the data sets. A firm's growth seems to depend on many interrelated factors whose quantification might still be a challenge for researchers who are interested in performing predictive analysis.

Besides identifying HGFs, other contributions attempt to maximize predictive power of future performance measures using sophisticated methods such as ANN or ensemble learners \citep[e.g.,][]{ravi2008soft,lam2004neural}. Even though these approaches achieve better results than traditional benchmarks, such as financial returns of market portfolios, a lot of variation of the performance measure is left unexplained. More importantly, the usage of such ``black-box'' tools makes it difficult to derive useful recommendations on what options exist to better individual firm performance. The fact that data sets and algorithm implementation are usually not made publicly available adds to our impotence at using such results as base for future investigations.

Yet, SL algorithms may help individual firms improve their performance from different perspectives. A good example in this respect is \citet{erel2018selecting} who showed how algorithms can contribute to appoint better directors. 
\label{subsec:performance}

\subsection{Financial distress and firm bankruptcy} \label{subsec:exit}

The estimation of default probabilities, financial distress and the predictions firms' bankruptcies based on balance sheet data and other sources of information on firms viability is a highly relevant topic for regulatory authorities, financial institutions and banks. In fact, regulatory agencies often evaluate the ability of banks to assess enterprises viability as this affects their capacity of best allocating financial resources and, in turn, their financial stability. Hence, the higher predictive power of SL algorithms can boost targeted financing policies that lead to safer allocation of credit either on the extensive margin, reducing the number borrowers by lending money just to the less risky ones, or on the intensive margin (i.e., credit granted), by setting a threshold to the amount of credit risk that banks are willing to accept.

In their seminal works in this field \cite{altman1968financial} and \cite{ohlson1980financial} apply standard econometric techniques such as multiple discriminant analysis (MDA) and logistic regression to assess the probability of firms' default. Moreover, since the Basel II Accord in 2004, default forecasting has been based on standard reduced-form regression approaches. However, these approaches may fail as for MDA the assumptions of linear separability and multivariate normality of the predictors may be unrealistic, and for regression models there may be pitfalls in (i) their ability to capture sudden changes in the state of the economy, (ii) their limited model complexity that rules out non-linear interactions between the predictors and, (iii) their narrow capacity for the inclusion of large sets of predictors due to possible multicollinearity issues. 

SL algorithms adjust for these shortcomings by providing flexible models that allow for non linear interactions in the predictors space and the inclusion of large number of predictors without the need to invert the covariance matrix of predictors, thus circumventing multicollinearity \citep{linn2019estimating}. Furthermore, as we saw in Section \ref{sec:sl}, SL models are directly optimized to perform predictive task and this leads, in many situations, to a superior predictive performance. 
In particular, \cite{moscatelli2019corporate} argue that SL outperform standard econometric models when the predictions of firms' distress is (i) based solely on financial accounts data as predictors and, (ii) relies on a large amount of data. In fact, as these algorithms are ``model-free", they need large data sets (``data hungry algorithms") in order to extract the amount of information needed to build precise predictive models.
Table \ref{tab:SL_bankcruptcy} depicts a number of papers in the field of economics, computer science, statistics, business and decision sciences that deal with the issue of predicting firms' bankruptcy or financial distress through SL algorithms. The former stream of literature (bankruptcy prediction) - that has its foundations in the seminal works of \cite{udo1993neural}, \cite{lee1996hybrid}, \cite{shin2005application} and \cite{chandra2009failure} - compares the binary predictions obtained with SL algorithms with the actual realized failure outcomes and uses this information to calibrate the predictive models. The latter stream of literature (financial distress prediction) - pioneered by \cite{fantazzini2009random} - deals with the problem of predicting default probabilities (DPs) \citep{moscatelli2019corporate, bargagli2020machine} or financial constraint scores \citep{linn2019estimating}. Even if these streams of literature approach the issue of firms' viability from slightly different perspectives, they train their models on dependent variables that range from firm's bankruptcy (see all the ``bankruptcy" papers in Table \ref{tab:SL_bankcruptcy}) to firm's insolvency \citep{bargagli2020machine}, default \citep{fantazzini2009random, behr2017default, moscatelli2019corporate}, liquidation \citep{bonello2018machine}, dissolvency \citep{bargagli2020machine} and financial constraint \citep{hansen2018predicting, sun2017dynamic}.

In order to perform these predictive tasks, models are built using a set of \textit{structured} and \textit{unstructured} predictors. With structured predictors we refer to balance sheet data and financial indicators, while unstructured predictors are, for instance, auditors reports, management statements and credit behaviour indicators. \cite{hansen2018predicting} show that the usage of unstructured data, in particular, auditors reports, can improve the performance of SL algorithms in predicting financial distress.
As SL algorithms do not suffer of multicollinearity issues, researchers can keep the set of predictors as large as possible. However, when researcher wish to incorporate just a set of ``meaningful" predictors, \cite{behr2017default} suggest to include indicators that (i) were found to be useful to predict bankruptcies in previous studies; (ii) are expected to have a predictive power based on firms' dynamics theory; (iii) were found to be important in practical applications.  As on the one side, informed choices of the predictors can boost the performance of the SL model, on the other side economic intuition can guide researchers in the choice of the best SL algorithm to be used with the disposable data sources. \cite{bargagli2020machine} show that a SL methodology that incorporates the information on missing data into its predictive model -- i.e., the BART-mia algorithm by \cite{kapelner2015prediction} -- can lead to staggering increases in the predictive performances when the predictors are missing-non-at-random (MNAR) and their missingness patterns are correlated with the outcome.\footnote{\cite{bargagli2020machine} argue that often times the decision not to release financial account information is driven by firm's financial distress.}

As different attributes can have different predictive power with respect to the chosen output variable, it may be the case that researchers are interested in providing to policy makers interpretable results in terms of which are the most important variables or the marginal effects of a certain variable on the predictions. Decision tree based algorithms, such as random forest \citep{breiman2001random}, survival random forests \citep{ishwaran2008random}, gradient boosted trees \citep{friedman2001greedy} and Bayesian additive regression trees \citep{chipman2010bart}, provide useful tool to investigate the aforementioned dimensions (i.e., variables importance, partial dependency plots, etc.). Hence, most of economics papers dealing with bankruptcy or financial distress predictions implement such techniques \citep{behr2017default, linn2019estimating, moscatelli2019corporate, bargagli2020machine} in service of policy relevant implications. On the other side, papers in the fields of computer science and business, that are mostly interested in the quality of predictions, de-emphasizing the interpretability of the methods, are built on black box methodologies such as artificial neural networks \citep{alaka2018systematic, bredart2014bankruptcy, hosaka2019bankruptcy, sun2011dynamic, tsai2008using, tsai2014comparative, wang2014improved, lee1996hybrid, udo1993neural}.
We want to highlight that, from the analyses of selected papers, we find no evidence of a positive correlation between the number of observations and predictors included in the model and the performance of the model. Indicating that the more is not always the better in SL applications to firm’s failures and bankruptcies.

\section{Final discussion} \label{sec:discussion} 
SL algorithms have advanced to become effective tools for prediction tasks relevant at different stages of the company life cycle. We provided a general introduction into the basics of SL methodologies and highlighted how they can be applied to improve predictions on future firm dynamics. 
In particular, SL methods improve over standard econometric tools in predicting firm success at an early stage, superior performance, and failure.
High dimensional, publicly available data sets have contributed in recent years to the applicability of SL methods in predicting early success on the firm level and, even more granular, success at the level of single products and projects.
While the dimension and content of data sets varies across applications, SVM and RF algorithms are oftentimes found to maximize predictive accuracy.
Even though the application of SL to predict superior firm performance in terms of returns and sales growth is still in its infancy, there is preliminary evidence that RF can outperform traditional regression-based models while preserving interpretability. Moreover, shrinkage methods, such as Lasso or stability selection, can help identifying the most important drivers of firm success. 
Coming to SL applications in the field of bankruptcy and distress prediction, decision tree-based algorithms and deep learning methodologies dominate the landscape, with the former widely used in economics due to their higher interpretability, and the latter more frequent in computer science where usually interpretability is de-emphasized in favour of higher predictive performance.

In general, the predictive ability of SL algorithms can play a fundamental role in boosting targeted policies at every stage of the lifespan of a firm - i.e., i) identifying projects and companies with a high success propensity can aid the allocation of investment resources; ii) potential high growth companies can be directly targeted with supportive measures; (iii) the higher ability to disentangle valuable and not valuable firms can act as screening device for potential lenders.

As granular data on the firm level becomes increasingly available, this will open many doors for future research directions focusing on SL applications for prediction tasks. To simplify future research in this matter, we briefly illustrated the principal SL algorithms employed in the literature of firm dynamics, namely decision trees, random forests, support vector machines and artificial neural networks. 
For a more detailed overview of these methods and their implementation in \texttt{R} we refer to our GitHub page \begin{footnotesize}(\url{github.com/fbargaglistoffi/machine-learning-firm-dynamics})\end{footnotesize} where we provide a simple tutorial to predict firms' bankruptcies. 

Besides reaching a high-predictive power, it is important, especially for policy makers, that SL methods deliver retractable and interpretable results. For instance, the US banking regulator has introduced the obligation for lenders to inform borrowers about the underlying factors that influenced their decision to do not provide access to credit.\footnote{These obligations were introduced by recent modification in the Equal Credit Opportunity Act (ECOA) and the Fair Credit Reporting Act (FCRA).}
Hence, we argue that different SL techniques should be evaluated, and researchers should opt for the most interpretable method when the predictive performance of competing algorithms is not too different.
This is central as the understanding of which are the most important predictors, or which is the marginal effect of a predictor on the output (e.g, via partial dependency plots) can provide useful insights for scholars and policy makers. 
Indeed, researchers and practitioners can enhance models' interpretability using a set of ready-to-use models and tools that are designed to provide useful insights on the SL black box. These tools can be grouped into three different categories: tools and models for (i) complexity and dimensionality reduction (i.e., variables selection and regularization via LASSO, ridge  or elastic net regressions, see \citet{martinez2011regularized}); (ii) model-agnostic variables' importance techniques (i.e., permutation feature importance based on how much the accuracy decreases when the variable is excluded, Shapley values, SHAP [SHapley Additive exPlanations], decrease in Gini impurity when a variable is chosen to split a node in tree based methodologies); and (iii) model-agnostic marginal effects estimation methodologies (average marginal effects, partial dependency plots, individual conditional expectations, accumulated local effects).\footnote{For a more extensive discussion on interpretability, models' simplicity and complexity we refer the reader to \cite{bargagli2020simplicity} and \cite{lee2020causal}.}

In order to form a solid knowledge base derived from SL applications, scholars should put an effort in making their research as replicable as possible in the spirit of Open Science.  Indeed, in the majority of papers that we analysed we did find no possibility to replicate the SL analyses. Higher standards of replicability should be reached by releasing details about the choice of the model hyperparameters, the codes and software used for the analyses as well as by releasing the training/testing data (to the extent that this is possible), anonymizing them in the case that the data are proprietary. Moreover, most of the datasets used for the SL analyses that we covered were not disclosed by the authors as they are linked to proprietary data sources collected by banks, financial institutions and business analytics firms (i.e., Bureau Van Dijk).

 Here, we want to stress once more time that SL learning per se is not informative about the causal relationships between the predictors and the outcome, so that researchers who wish to draw causal inference should carefully check the standard identification assumptions \citep{imbens2015causal} and inspect whether or not they hold in the scenario at hand \citep{athey2019machine}. Besides not directly providing causal estimands, most of the reviewed SL applications focus on pointwise predictions where inference is de-emphasized. Providing a measure of uncertainty about the predictions, e.g., via confidence intervals, and assessing how sensitive predictions are to unobserved predictors are sensible directions to explore further \citep{bargagli2020assessing}.

In this contribution, we focus on the analysis of how SL algorithms predict various firm dynamics on ``inter company data'' that cover information across firms. Yet, nowadays companies themselves apply ML algorithms for various clustering and predictive tasks \citep{lee2019machine}, which will presumably  become more prominent for small and medium sized companies (SMEs) in the upcoming years. This is due to the fact that (i) SMEs start to construct proprietary data bases (ii) develop the skills to perform in-house ML analysis on this data and (iii) powerful methods are easily implemented using common statistical software.

Against this background, we want to stress that applying SL algorithms and economic intuition regarding the research question at hand should ideally complement each other. Economic intuition can aid the choice of the algorithm and the selection of relevant attributes, thus leading to better predictive performance \citep{bargagli2020machine}. Further, it requires a deep knowledge of the studied research question to properly interpret SL results and to direct their purpose so that \textit{intelligent machines are driven by expert human beings}.

\begin{landscape}
  \begin{table}[p]
    \begin{minipage}[t][\textheight][t]{\linewidth}
      \caption{SL literature on firms' early success and innovation}
      \label{tab:SL_earlysuccess}
      \begin{footnotesize}
       \renewcommand{\arraystretch}{1.1}
        \begin{tabularx}{\textwidth}{>{\arraybackslash}p{4.1cm}>{\arraybackslash}p{1.5cm}>{\arraybackslash}p{2.9cm}>{\arraybackslash}p{2.9cm}>{\arraybackslash}p{1.6cm}>{\arraybackslash}p{1.9cm}>{\arraybackslash}p{1.4cm}>{\arraybackslash}p{1.9cm}}
        \hline\noalign{\smallskip}
           \textbf{Author, Year} & \textbf{Domain} &\textbf{Output} &\textbf{Country, time} & \textbf{Data set size} &  \textbf{Primary \newline SL-method} & \textbf{Attributes} &  \textbf{GOF}  \tabularnewline
\noalign{\smallskip}\svhline\noalign{\smallskip}
\cite{arroyo2019assessment}& CS & Startup funding & INT (2011-2018)& 120,507 & GTB & 105 & 82\% (ACC)
\tabularnewline
\cite{bento2018predicting} & BI & Startup funding & USA (1985-2014) & 143,348 & RF  &  158 & 93\% (AUC)
\tabularnewline
 \cite{bohm2017business}& BI  &Startup survival,\newline -growth & USA, GER (1999-2015) & 181 & SVM & 69 &67-84\% (ACC) 
\tabularnewline
\cite{guerzoni2019survival} & ECON & Startup \newline innovativeness & ITA (2013) & 45,576 & bagging, ANN & 262 & 56\% (TPR),\newline 95\% (TNR) 
\tabularnewline
\cite{kinne2019predicting} & ECON & Firm innovativeness & GER (2012-2016) & 4,481& ANN & N/A & 80\% (F-score)  
\tabularnewline
\cite{krishna2016predicting} & CS & Startup survival & INT (1999-2014) & 13,000 & RF, LR  & 70 & 73-96\% (ACC) 
\tabularnewline
\cite{mckenzie2017man} & ECON & Startup survival & NIG (2014-2015) & 2,506 & SVM & 393 & 64\% (ACC)
\tabularnewline
\cite{sharchilev2018web} & CS &Startup funding & INT  & 21,947 & GTB & 49 & 85\% (AUC)
\tabularnewline
\cite{xiang2012supervised} & BI & Startup M\&A &  INT (1970-2007) & 59,631 &  BN & 27 & 68-89\% (AUC) 
\tabularnewline
\cite{yankov2014models} & ECON & Startup survival & BUL  & 142 & DT & 15 & 67\% (ACC)
\tabularnewline
\cite{zhang2017predicting} & CS & Startup funding & INT (2015-2016) & 4,001 & SVM & 14 & 84\% (AM) 
\tabularnewline
\noalign{\smallskip}\hline\noalign{\smallskip}
  \cite{dimasi2015tool} & PHARM & Project success \newline (oncology drugs)  & INT (1999-2007) & 98 & RF & 4 & 92\% (AUC) 
  \tabularnewline
  \cite{etter2013launch} & CS &  Project funding & INT (2012-2013) & 16,042 & Ensemble SVM & 12& $>$ 76\% (ACC) 
  \tabularnewline
   \cite{feijoo2020key} & PHARM & Project success \newline (clinical trials)  & INT (1993-2018) & 6,417 & RF &  17 & 80\% (ACC) 
   \tabularnewline
\cite{greenberg2013crowdfunding} & CS & Project funding & INT (2012) & 13,000 & RF & 12 & 67\% (ACC) 
\tabularnewline
\cite{kaminski2019predicting} & ECON & Project funding & INT (2009-2017) & 20,188 & LR & 200 & 65-71\% (ACC) 
\tabularnewline
 \cite{kyebambe2017forecasting} & BMA &Emerging Technologies & USA (1979 - 2010) & 11,000 & SVM & 7 & 71\% (ACC)
\tabularnewline
 \cite{lo2019machine} &CS & Project success (drugs)   & INT (2003-2015) & 27,800 & KNN,RF  & 140 & 74-81\% (AUC) 
 \tabularnewline
\cite{munos2019}  & PHARM & Project success (drugs)    & USA (2008-2018) & 8.800 & BART  & 37 & 91-96\% (AUC)  
\tabularnewline
\cite{rouhani2013erp} & ENG & Project success (IT) & ME (2011) & 171 & ANN & 24 & 69\% (ACC) \tabularnewline
\noalign{\smallskip}\hline\noalign{\smallskip}
\end{tabularx}

\end{footnotesize}
     \begin{scriptsize}
      Abbreviations used - Domain: ECON: Economics, CS: Computer Science, BI: Business Informatics, ENG: Engineering, BMA: Business, Management and Accounting, PHARM: Pharmacology. Country: ITA: Italy, GER: Germany, INT: International, BUL: Bulgaria, USA: United states of America, NIG: Nigeria, ME: Middle East. Primary SL-method: ANN: (deep) neural network, SL: supervised learner, GTB: gradient tree boosting, DT: Decision Tree, SVM: support vector machine, BN: Bayesian Network, IXL: induction on eXtremely Large databases, RF: random forest, KNN: k-nearest neighbour, BART: Bayesian additive regression tree, LR: Logistic regression, TPR: true positive rate,  TNR: true negative rate, ACC: Accuracy, AUC: Area under the receiver operating curve, BACC: Balanced Accuracy (average between TPR and TNR). The year was not reported when it was not possible to recover this information from the papers.
       \end{scriptsize}  
    \end{minipage}
  \end{table}
\end{landscape}

 \begin{landscape}
  \begin{table}[p]
    \begin{minipage}[t][\textheight][t]{\linewidth}
      \caption{SL literature on firms' growth and performance}
      \label{tab:SL_performance}
        \renewcommand{\arraystretch}{1.1}
        \begin{tabularx}{\textwidth}{>{\arraybackslash}p{4.1cm}>{\arraybackslash}p{1.5cm}>{\arraybackslash}p{2.9cm}>{\arraybackslash}p{2.9cm}>{\arraybackslash}p{1.6cm}>{\arraybackslash}p{1.9cm}>{\arraybackslash}p{1.4cm}>{\arraybackslash}p{1.9cm}}
        \hline\noalign{\smallskip}
           \textbf{Author, Year} & \textbf{Domain} &\textbf{Output} &\textbf{Country, time} & \textbf{Data set size} &  \textbf{Primary \newline SL-method} & \textbf{Attributes} &  \textbf{GOF}  \tabularnewline
\noalign{\smallskip}\svhline\noalign{\smallskip}
\cite{weinblat2018forecasting} & BMA & High growth firms &  INT (2004-2014) & 179,970 & RF & 30 & 52\%-81\% (AUC) \\
\tabularnewline
\cite{megaravalli2019predicting} & BMA & High growth firms & ITA (2010-2014) & 22,333  & PR* & 5 & 71\% (AUC) \\ 
\tabularnewline
\cite{coad2019catching} & BMA & High growth firms &  HRV (2003-2016) & 79,109 & Lasso &  172 & 76\% (ACC) \\
\cite{miyakawa2017forecasting} & ECON & Firm exit, sales growth, profit growth & JPN   (2006-2014) & 1,700,000 & weighted RF & ~50 & 70\%,68\%,61\% (AUC)\\
\tabularnewline
\cite{lam2004neural} & BI & ROE & USA (1985-1995) & 364 firms per set &  ANN & 27 &  Portfolio return comparison\\
\tabularnewline
\cite{kolkman2019data} & ECON & Asset growth & NL & 8,163 firms & RF & 113 & 16\% ($R^2$)\\
\tabularnewline
\cite{qiu2014supervised} & CS & Groups of SAR,\newline EPS growth  & USA (1997-2003) & 1,276 firms & SVM & From annual reports &~50\% (ACC) \\
\tabularnewline
\cite{bakar2009applying} & BMA & ROA & MYS (2001-2006) & 91 & ANN & 7 & 66.9\% ($R^2$)\\
\tabularnewline
\cite{baumann2015maximize} & CS & Customer Churn & INT & 5000-93,893 & Ensemble & 20-359 & 1.5 - 6.8 $(L_1$) \\
\tabularnewline
\cite{ravi2008soft} & CS & Profit of banks & INT (1991-1993) &  1000 & Ensemble & 54 & 80-93\% (ACC)\\
\tabularnewline

\noalign{\smallskip}\hline\noalign{\smallskip}
 
    \end{tabularx}
     
     \begin{scriptsize}
      Abbreviations used - Domain: ECON: Economics, CS: Computer Science, BI: Business Informatics, BMA: Business, Management and Accounting. Country: ITA: Italy, INT: International, HRV: Croatia, USA: United states of America, JPN: Japan, NL: Netherlands, MYS: Malaysia. Primary SL-method: ANN: (deep) neural network, SVM: support vector machine, RF: random forest, PR: Probit regression (simplest form of SL if out of sample performance analysis used), Lasso: Least absolute shrinkage and selection operator, Ensemble: Ensemble Learner. GOF: Accuracy, AUC: Area under the receiver operating curve, $L_1$: Top decile lift. $R^2$ R-squared. The year was not reported when it was not possible to recover this information from the papers. 
       \end{scriptsize} 
    \end{minipage}
  \end{table}
\end{landscape}

\begin{landscape}
  \begin{table}[p]
    \begin{minipage}[t][\textheight][t]{\linewidth}
      \caption{SL literature on firms' failure and financial distress}
      \label{tab:SL_bankcruptcy}
        \begin{tabular}{p{3.5cm}p{1.7cm}p{2.2cm}p{2.9cm}p{1.5cm}p{2.9cm}p{1.5cm}p{2.0cm}}
        \hline\noalign{\smallskip}
           \textbf{Author, Year} & \textbf{Domain} &\textbf{Output} &\textbf{Country, time} & \textbf{Data set size} &  \textbf{Primary SL-method} & \textbf{Attributes} &  \textbf{GOF} \tabularnewline
\noalign{\smallskip}\svhline\noalign{\smallskip} 
  \cite{alaka2018systematic} & CS & Bankruptcy & UK (2001-2015) & 30,000 & NN & 5 & 88\% (AUC) \tabularnewline
  \cite{barboza2017machine} & CS & Bankruptcy & USA (1985-2014) & 10,000 & SVM, RF, BO, BA & 11 &  93\% (AUC)  \tabularnewline
  \cite{bargagli2020machine} & ECON & Fin. distress & ITA (2008-2017) & 304,000 & BART & 46 & 97\% (AUC) \tabularnewline
  \cite{behr2017default} & ECON & Bankruptcy & INT (2010-2011) & 945,062 & DT, RF & 20 &  85\% (AUC) \tabularnewline
  \cite{bonello2018machine} & ECON & Fin. distress & USA (1996-2016) & 1,848 & NB, DT, NN & 96 & 78\% (ACC) \\
  \cite{bredart2014bankruptcy} & BMA & Bankruptcy & BEL (2002-2012) & 3,728 & NN & 3 & 81\%(ACC) \\
    \cite{chandra2009failure} & CS & Bankruptcy & USA (2000) & 240 & DT & 24 &  75\%(ACC) \\
  \cite{cleofas2016financial} & CS & Fin. distress & INT (2007) & 240-8,200 & SVM, NN, LR & 12-30 & 78\% (ACC)  \tabularnewline
  \cite{danenas2015selection} & CS & Fin. distress & USA (1999-2007) & 21,487 & SVM, NN, LR & 51 & 93\% (ACC)  \tabularnewline
  \cite{fantazzini2009random} & STAT & Fin. distress & DEU (1996-2004) & 1,003 & SRF & 16 & 93\% (ACC) \tabularnewline
  \cite{hansen2018predicting} & ECON & Fin. distress & DNK (2013-2016) & 278,047 & CNN, RNN & 50 &  84\% (AUC) \tabularnewline
  \cite{heo2014adaboost} & CS  & Bankruptcy & KOR (2008-2012) & 30,000 & ADA & 12 & 94\% (ACC)  \tabularnewline
  \cite{hosaka2019bankruptcy} & CS & Bankruptcy & JPN (2002-2016) & 2,703 & CNN & 14  & 18\% (F-score)  \tabularnewline
  \cite{kim2014predicting}  & CS & Bankruptcy & KOR (1988-2010) & 10,000 & ADA, DT & 30 & 95\% (ACC) \tabularnewline
  \cite{lee1996hybrid} & BMA  & Bankruptcy & KOR (1979-1992) & 166 & NN & 57 & 82\% (ACC) \tabularnewline
  \cite{liang2016financial} & ECON & Bankruptcy & TWN (1999-2009) & 480 & SVM, KNN, DT, NB & 190 & 82\% (ACC) \tabularnewline
  \cite{linn2019estimating} & ECON & Fin. distress &  INT (1997-2015) &  48,512 & DRF & 16 & 15\% ($R^2$) \tabularnewline
  \cite{moscatelli2019corporate} & ECON & Fin. distress & ITA (2011-2017) & 250,000 & RF & 24 & 84\%(AUC)
  \tabularnewline
  \cite{shin2005application} & CS & Bankruptcy & KOR (1996-1999) &  1,160 & SVM & 52 & 77\%(ACC) \tabularnewline
  \cite{sun2011dynamic} & CS & Bankruptcy &  CHN  & 270 & CBR, KNN & 5 & 79\% (ACC) \tabularnewline
  \cite{sun2017dynamic} & BMA & Fin. distress & CHN (2005-2012) & 932 & ADA, SVM & 13 & 87\%(ACC) \tabularnewline
  \cite{tsai2008using} & CS & Bankruptcy & INT & 690-1,000 & NN & 14-20 &  79-97\%(ACC) \tabularnewline
  \cite{tsai2014comparative} & CS & Bankruptcy & TWN & 440 & ANN, SVM, BO, BA & 95 & 86\% (ACC) \tabularnewline
  \cite{wang2014improved} & CS & Bankruptcy & POL (1997-2001) & 240 & DT, NN, NB, SVM & 30 & 82\% (ACC) \tabularnewline
   \cite{udo1993neural} & CS & Bankruptcy & KOR (1996-2016) & 300 & NN & 16 & 91\% (ACC) \tabularnewline
  \cite{zikeba2016ensemble} & CS & Bankruptcy & POL (2000-2013) & 10,700 & BO &  64 & 95\% (AUC)
  \tabularnewline
  \noalign{\smallskip}\hline\noalign{\smallskip}
\end{tabular}%

  \begin{scriptsize}
      Abbreviations used - Domain: ECON: Economics, CS: Computer Science, BMA: Business, Management, Accounting, STAT: Statistics. Country: BEL: Belgium, ITA: Italy, DEU: Germany, INT: International, KOR: Korea, USA: United states of America, TWN: Taiwan, CHN: China, UK: United Kingdom, POL: Poland. Primary SL-method: ADA: AdaBoost, ANN: Artificial neural network, CNN: Convolutional neural network, NN: Neural network, GTB: gradient tree boosting, RF: Random forest, DRF: Decision random forest, SRF: Survival random forest, DT: Decision Tree, SVM: support vector machine, NB: Naive Bayes, BO: Boosting, BA: Bagging, KNN: k-nearest neighbour, BART: Bayesian additive regression tree, DT: decision tree, LR: Logistic regression. Rate: ACC: Accuracy, AUC: Area under the receiver operating curve.
      The year was not reported when it was not possible to recover this information from the papers. 
       \end{scriptsize}  
    \end{minipage}
  \end{table}
\end{landscape}

	\bibliographystyle{apalike}
	\bibliography{references}
	\pagebreak
	

\end{document}